\documentclass[11pt, conference, onecolumn]{IEEEtran}
\usepackage[margin=1in]{geometry} 
\IEEEoverridecommandlockouts
\usepackage{setspace}
\usepackage{comment} 
\usepackage{amsmath,amssymb,amsfonts}
\usepackage{algorithmic}
\usepackage{graphicx}
\usepackage{textcomp}
\usepackage{xcolor}
\usepackage{hyperref}
\usepackage{biblatex} 
\makeatletter
\newcommand{\linebreakand}{%
  \end{@IEEEauthorhalign}
  \hfill\mbox{}\par
  \mbox{}\hfill\begin{@IEEEauthorhalign}
}
\makeatother

\begin{document}

\title{Survey of Security Issues in Memristor-based Machine Learning Accelerators for RF Analysis}

\author{
    \IEEEauthorblockN{William Lillis}
    \IEEEauthorblockA{\textit{Electrical And Computer Engineering}\\
    \textit{University of Massachusetts Amherst}\\
    wlillis@umass.edu}
\and
    \IEEEauthorblockN{Max Cohen Hoffing}
    \IEEEauthorblockA{\textit{Electrical And Computer Engineering}\\
    \textit{University of Massachusetts Amherst}\\
    mcohenhoffin@umass.edu}
\linebreakand
    \IEEEauthorblockN{Wayne Burleson}
    \IEEEauthorblockA{\textit{Electrical And Computer Engineering}\\
    \textit{University of Massachusetts Amherst}\\
    burleson@umass.edu}
}

\maketitle
\thispagestyle{plain}
\pagestyle{plain}

\begin{abstract}
We explore security aspects of a new computing paradigm that combines novel memristors and traditional Complimentary Metal Oxide Semiconductor (CMOS) to construct a highly efficient analog and/or digital fabric that is especially well-suited to Machine Learning (ML) inference processors for Radio Frequency (RF) signals.  
Memristors have different properties than traditional CMOS which can potentially be exploited by attackers.  In addition, the mixed signal approximate computing model has different vulnerabilities than traditional digital implementations.  However both the memristor and the ML computation can be leveraged to create security mechanisms and countermeasures ranging from lightweight cryptography, identifiers (e.g. Physically Unclonable Functions (PUFs), fingerprints, and watermarks), entropy sources, hardware obfuscation and leakage/attack detection methods. 
Three different threat models are proposed: 1) Supply Chain, 2) Physical Attacks, and 3) Remote Attacks.  For each threat model, potential vulnerabilities and defenses are identified.
This survey reviews a variety of recent work from the hardware and ML security literature and proposes open problems for both attack and defense.  The survey emphasizes the growing area of RF signal analysis and identification in terms of the commercial space, as well as military applications and threat models. We differ from other other recent surveys \cite{zou_review_2022} that target ML in general, neglecting RF applications. 

\end{abstract}


\thanks{
{
\begin{spacing}{0.9}
    \footnotesize This research was sponsored by the Army Research Laboratory through the Kostas Research Institute and Mass Nanotech and was accomplished under Cooperative Agreement Number W911NF-23-2-0014. The views and conclusions contained in this document are those of the authors and should not be interpreted as representing the official policies, either expressed or implied, of the Army Research Laboratory or the U.S. Government. The U.S. Government is authorized to reproduce and distribute reprints for Government purposes notwithstanding any copyright notation herein.
\end{spacing}
}
}

\section{Introduction}

With the recent impact of ML in fields such as Robotics, Autonomous Vehicles, Natural Language Processing, and RF analysis among others, practical ML technologies are developing at an astonishing rate. Encouraged by a long list of successful applications in a wide variety of fields, investment in ML technologies (in particular various Neural Network architectures) is expected to continue its explosive growth for the foreseeable future. As ML technology becomes pervasive and increasingly integrated into our lives, the safety and protection of critical information in ML systems is paramount. Security analysis and development of countermeasures should to be performed in a holistic manner, since many attacks transcend layers of computation, spanning from algorithms to hardware. 

Memristors are an emergent nanoelectric technology that offer numerous benefits when deployed in large arrays, including acceleration and efficiency gains in Neural Network (NN) computation. Memristor-based accelerators have demonstrated particular effectiveness in Edge-deployed NN inference roles due to their speed and power efficiency \cite{wang_resistive_2020}. 

As integration technology matures, memristors will likely be configured in hybrid chiplet-based systems that combine standard digital CMOS (typically programmable Central Processing Units (CPUs), Graphical Processing Units (GPUs), Field Programmable Gate Arrays (FPGAs), Application Specific Integrated Circuits (ASICS)), volatile and non-volatile memory technologies with analog sensing front-ends.  Trends indicate increasing investment into chiplet architectures, and this will likely be reflected in memristor technology moving forward \cite{heyman_march_2022, clark_us_2023}.  Memristors, however, raise several new concerns and capabilities for security; including some specific to ML computation, the overall deployment platform and application area \cite{zeitouni_security_2020}.  

This survey emphasizes the growing area of RF signal analysis and identification in terms of the commercial space, as well as threat models for applications in law enforcement and the military. This contrasts with other recent security surveys of memristors for ML \cite{zou_review_2022} that target ML in general.  RF signal analysis and identification differs from other ML applications in several ways:

\ 

\begin{enumerate}
    \item Data-rates from an RF front-end can be very high and involve numerous channels.
    \item The number of different devices, signals and protocols to be identified can vary.
    \item Low-latency computation is required to enable fast real time response, which can be critical for both legitimate applications (e.g. Cognitive radio) as well as to identify and defend against active attacks.
    \item Application scenarios can involve several different parties producing and analyzing signals with different threat models.
\end{enumerate}

\ 

Our goal is to present threat models for Memristor-based ML Edge computing for RF Analysis under three different scenarios: Supply Chain Attacks, Physical attacks and Remote attacks. Combinations of the three threat models are also considered. The standard procedure in threat modeling is to identify system vulnerabilities, define security requirements, pinpoint capabilities and motives of the attacker and then devise potential countermeasures and defenses \cite{anderson_security_2020}. While we cover a wide range of vulnerabilities, attacks, and defenses, our survey is not an all-inclusive survey and is intentionally tailored to the RF signal analysis space. Threats and security capabilities that involve multiple levels of abstractions are investigated (e.g. hardware, software, algorithm and application).

We proceed in the following sequence: In Section II we provide background on memristors as used in ML computation. In Section III we discuss various ML accelerators, focusing on key differences with more traditional computing models. In Section IV we consider the applications to Edge Computing, in particular for RF Signature Analysis. Section V introduces and provides a description of the three threat models: 1) Supply Chain Attacks, 2) Physical Attack and 3) Remote Attack. For each threat model we identify potential vulnerabilities and defenses. Section VI discusses potential security countermeasures for memristor architectures. Section VII describes handling of anomalies and faults in memristor nodes. Section VIII discusses strategies for integrating memristors with more traditional CMOS technology, as well as the associated additional security concerns and defenses caused by such integration. Our conclusions are contained in Section IX.

\section{Background on Memristor Technology}

Memristors are an emergent device technology, primarily considered as a Non-Volatile memory device. The device was initially proposed by Leon Chua in 1971 as the fourth fundamental circuit element which is ``as basic as the three classical circuit elements already in existence, namely, the resistor, inductor, and capacitor" \cite{chua_memristor-missing_1971}. Memristors are typically constructed with resistive switching materials (RSMs) which exhibit hysteresis. RSMs currently utilize one of four physical mechanisms to achieve hysteresis: redox reactions, phase transitions, spin-polarized tunnelling and ferroelectric polarization. Each mechanism provides a set of trade offs, with different values for switching speed, switching energy, physical size, data retention and endurance, conductance range, number of distinguishable conductance states, current-voltage linearity, and many other relevant properties \cite{wang_resistive_2020}.

Due to the increasing strain in sustainment of Moore's law reflected in current CMOS devices and circuitry, the rise in popularity of data-intensive tasks such as ML, and the limits imposed by the ``von-Neumann bottleneck", in certain contexts in-memory computing presents itself as an attractive alternative to more traditional schemes \cite{wang_resistive_2020, xia_memristive_2019}. Traditionally expensive operations such as vector matrix multiplication (VMM) can be completed in $\mathcal{O}(1)$ time by leveraging physical laws; namely Kirchhoff's Law for summation and Ohm's Law for multiplication.  Recent advances are now demonstrating the capabilities of these devices, including wireless communications \cite{wang_parallel_2023}, fully analog ML accelerators \cite{kiani_fully_2021} and greatly expanded storage density \cite{rao_memristor_2023}.

\begin{figure}
    \centering
    \includegraphics[scale=0.3]{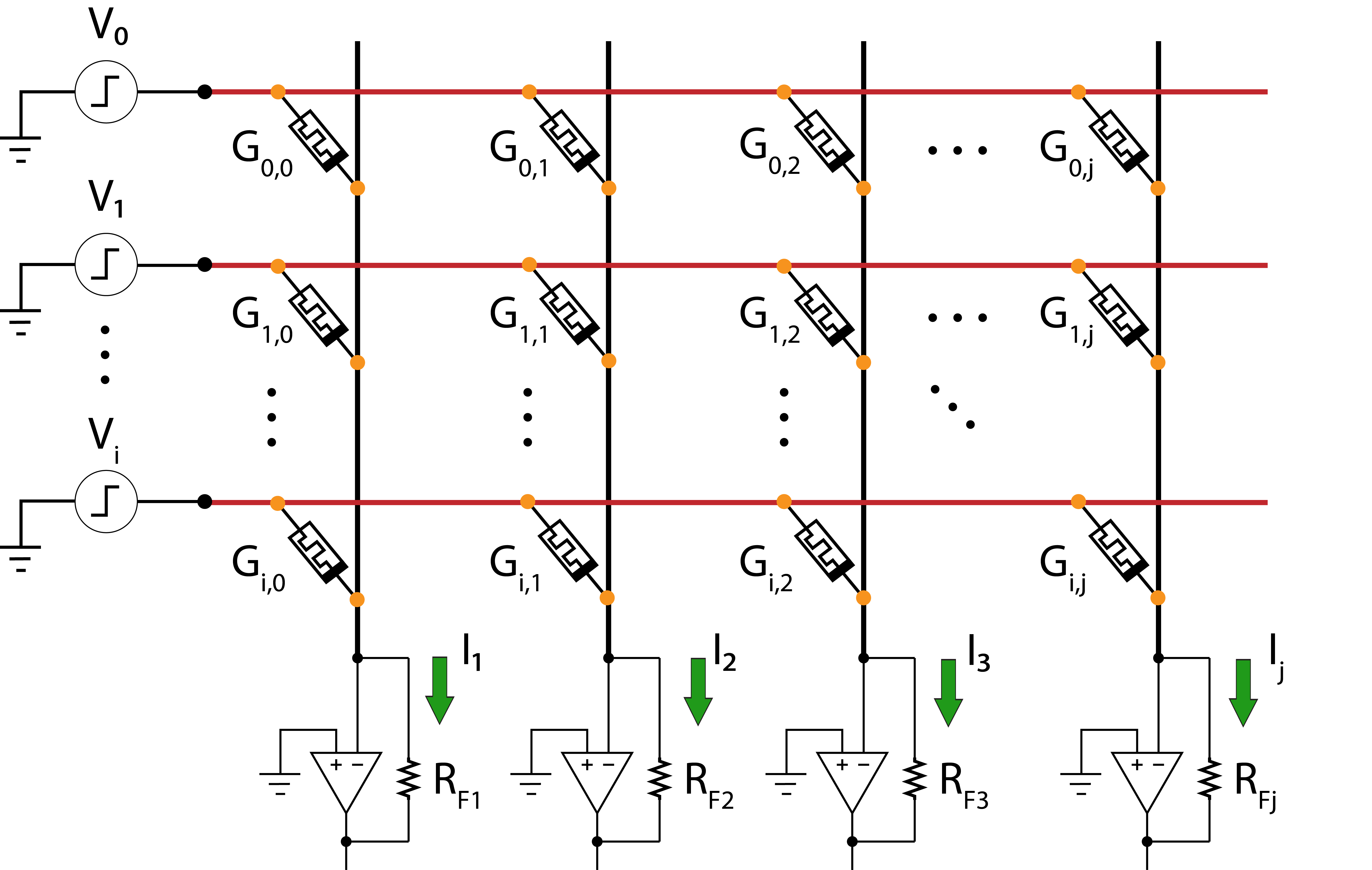}
    \caption{An example of a memristor array. This array shows the basics of a current summation model. The voltage inputs are multiplied against the conductances ($G_{i,j}$) of each device which generates current. The currents are then accumulated at the output.}
    \label{fig:Memristor_Array}
\end{figure}

Given this set of unique properties and capabilities, memristors are well suited for ML acceleration roles. In particular, memristive crossbars have been successfully deployed in NN training and inference roles \cite{li_efficient_2018}. Figure 1 shows the basic principles of a memristive crossbar array used to perform VMM in an analog domain. 

\section{Background on ML Accelerators}

Deep Learning has recently made great strides in a wide variety of applications such as object recognition and natural language processing \cite{lecun_deep_2015}. Along with the increased expressiveness and capabilities of these models, however, comes a rising need for more powerful yet energy-efficient processors, especially in Internet of Things (IoT) and Edge applications. In response, new hardware (e.g. traditional and more recently enterprise GPUs) has been developed to optimize the training and usage of these models. However, these accelerators are still limited by the von-Neumann architecture; hampered by ``the constant data shuttling between the information processing and memory units. In addition, the performance mismatch between the processor and memory units leads to considerable latency (referred to as the `memory wall')" \cite{xia_memristive_2019}. In response to these challenges, increased attention has been given to \textit{in situ}, or \textit{in memory} computing: an alternative schema in which memory is used for both storage and computation \cite{xiao_accuracy_2022}. In particular, memristive crossbar arrays have been successfully used as in-memory analog ML accelerators \cite{xia_memristive_2019, li_efficient_2018, kiani_fully_2021}.

\subsection*{Approximate Computing}

The \textit{in situ} computing offered by memristive crossbar arrays is a form of approximate computing, in which a shift is made ``from conventional precise processing to inexact computation but still satisfying the system requirement on accuracy" \cite{yellu_security_2021}. Memristor computation falls into this category due to the inherently inexact nature of analog compared to classical digital computation. While this new paradigm offers performance improvements, it also exposes several new attack surfaces. Yellu et al. \cite{yellu_security_2021} introduces four new potential attack scenarios: Building Covert Channel (BCC), Error Compensation (ECA), Tampering Error Resilience Mechanism (TERM) and Error Propagation (ECA) attacks. BCC attacks involve the exploitation of approximate compute modules in order to open passive side channels to enable leakage of covert information. ECA attacks exploit the imprecise nature inherent to approximate computing to insert stealthy malicious behavior.  TERM attacks are inserted into and around the system's Error Monitor, Precision Cutoff Threshold, and Accuracy Tuner, generally for the purpose of denial-of service or reverse engineering. Finally, EPA attacks allow the adversary to utilize unsecured approximate-computing modules to forward errors through the rest of the data path. The authors stress the need for new accuracy metrics to be developed in order to differentiate benign from malicious errors. Further research is required to investigate these potential threats and countermeasures in a more concrete manner.

\section{RF Signature Analysis} \label{sec:RF Analysis}

As ML (and in particular, deep learning) becomes ever-more ubiquitous \cite{lecun_deep_2015}, the same is true for the RF analysis domain. In particular, ``deep feature learners with an inherent recurrent structure have been shown to perform well" for RF ML tasks \cite{roy_machine_2019}. Certain ML architectures have demonstrated strong performance in analyzing different RF signal features. Deep Neural Networks (DNNs) have performed well for fixed value parameters such as rise time, Convolutional Neural Networks (CNNs) have performed well for spatially correlated data such as modulation techniques and Recurrent Neural Networks (RNNs) have performed well with temporally correlated data such as I/Q signals \cite{roy_machine_2019}. LeCun et al. argues that deep learning is a technique that ``allows a machine to be fed with large amounts of raw data and to automatically discover the representations needed for detection or classification" \cite{lecun_deep_2015}. As with other challenging application domains, the rich expressiveness of deep learning models has allowed success where other ML architectures have fallen short.  However, the effective application of ML to the RF space is still a non-trivial problem. The inherently noisy nature of RF communications, non-trivial variation in environmental conditions caused by the ‘train on one day test on another day’ problem \cite{mohanti_airid_2020}, lossy conversions between the analog and digital domains, and other issues complicate matters greatly. Extensive dataset and feature engineering (e.g. adding noise, selectively removing parts of the signal, or factoring in ``significant modeling insights" \cite{cekic_wireless_2021}) has continued to prove necessary to achieve acceptable performance in recent work. Various commercializations of this technology target lucrative markets such as spectrum-awareness \cite{inc_spectrum_nodate}.

One relevant application of RF signature analysis is wireless fingerprinting. This technique aims to exploit the hardware variations in analog circuitry, which are then reflected in the signals transmitted by said hardware. These variations can enable unique identification of different wireless transmitters, regardless of the actual data content of their transmission. The ability to differentiate between multiple (potentially adversarial) transmitters is a powerful physical layer technique, allowing one to detect imposters even after security protocols farther up the network stack have been compromised. Figure 2 illustrates the basic concept of RF Signal identification.  Different approaches have utilized both the transient and steady state portion of the signal, predefined (e.g. modulation errors) and inferred features (e.g. information extracted as a result of applying a Fast Fourier Transform (FFT)), as well as a single or multiple features. Ideally a wireless fingerprint is robust to variations such as environmental conditions, voltage and power levels, relative orientation, and physical distance between transmitter and receiver \cite{danev_physical-layer_2012}. The use of complex-valued DNNs in combination with various feature ``augmentation strategies" has shown promise in developing protocol agnostic and variation-resilient fingerprinting techniques \cite{gopalakrishnan_robust_2019, kokalj-filipovic_adversarial_2019, cekic_wireless_2021}.

\begin{figure}
    \centering
    \includegraphics[scale=0.3]{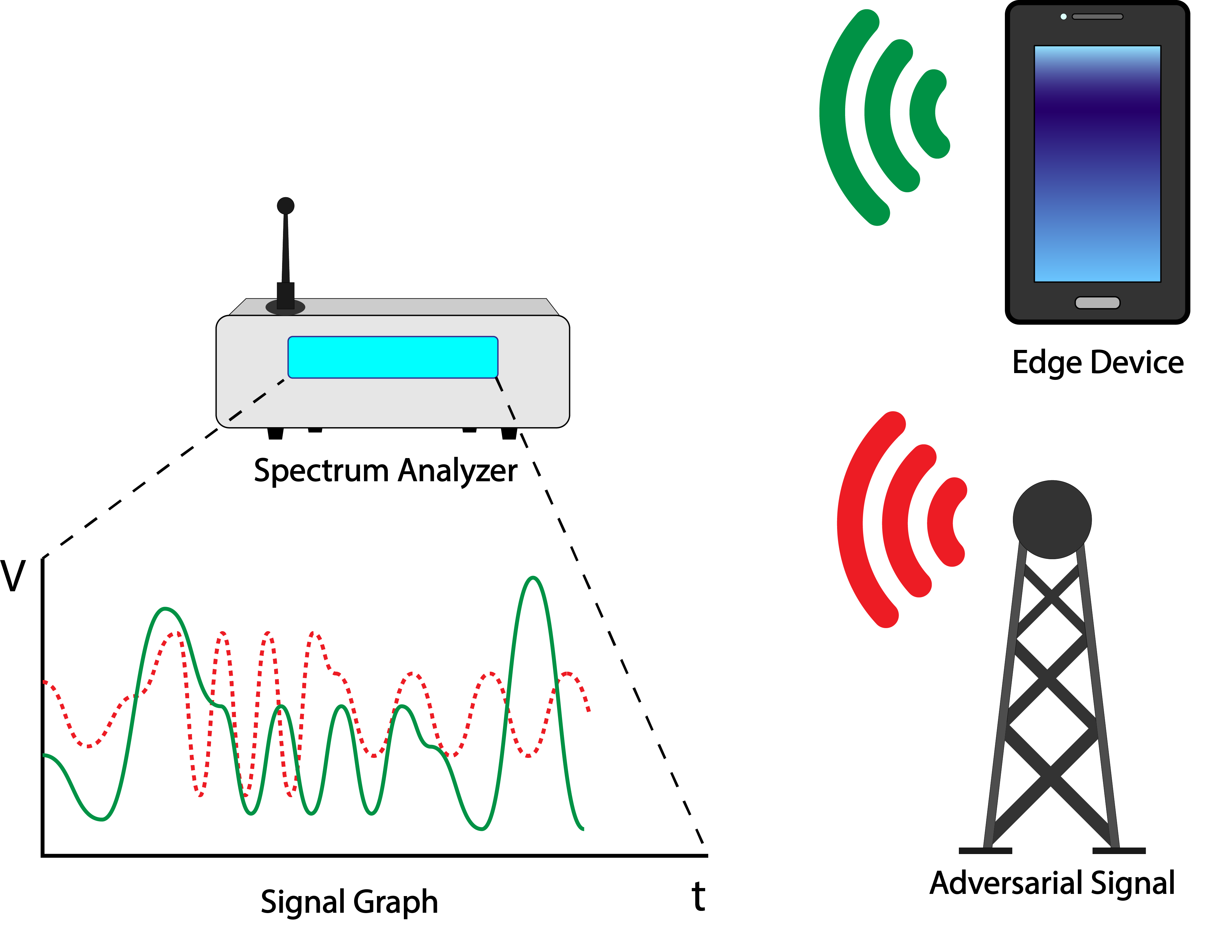}
    \caption{An example of a spectrum analyzer differentiating between transmitters sending a friendly signal and an adversarial signal based on inherent variation in transmitter properties.}
    \label{fig:Signal_Classifier}
\end{figure}

Autoencoder pre-training has been shown to mitigate the effectiveness of adversarial examples in the RF space. CNNs have also shown promise in the application of wireless transmitter fingerprinting. Complex-valued weights have shown advantages over their real-valued counterparts in these tasks, since RF data is commonly represented in a complex form. As demonstrated by Wang et al., memristor technology is fully capable of processing complex-valued data and takes advantage of these specialized architectures \cite{wang_parallel_2023}. Difficulties still exist in the elimination of confounding factors. In addition, the difficulties of adversarial require a more thorough exploration \cite{gopalakrishnan_robust_2019, cekic_wireless_2021}.

In addition to the wireless fingerprinting of one or several wireless entities, it is also worthwhile to take a broader view of the wireless medium a device is operating within. Cognitive radio allows more efficient use of the RF spectrum by dynamically adapting parameters to the current state of a given device's wireless environment. Indeed, such RF networks are ``extremely agile, allowing communications to move around the frequency domain optimizing capacity when users are present. These systems require components that can quickly and accurately detect whether or not signals are present in their current channel and what channels are free around them" \cite{inc_spectrum_nodate}. More efficient use of the shared wireless medium will become increasingly important as the number of wireless devices continues to increase \cite{muchandi_cognitive_2016}.

Furthermore, memristive crossbars show great promise in non-ML RF applications. By connecting memristive crossbars directly with analog wireless media, Wang et al. was able to ``store, transfer and modulate time-varying analogue signals and parallelly process data without digital-to-analogue converters (DACs) and analogue-to-digital converters (ADCs)" \cite{wang_parallel_2023}. This integration yielded an increase in power efficiency of over $100$ fold. Zen et al. created a CMOS-memristor hybrid baseband processor, ``implementing a full-fledged communication system" that outperforms its CMOS-only counterparts in speed and energy efficiency by factors of $10^3$ and $10^6$ respectively \cite{zeng_realizing_nodate}. The recent progress in implementing various ML architectures via memristive crossbar arrays \cite{kiani_fully_2021, li_efficient_2018, xia_memristive_2019} shows promise for more tightly integrated RF Machine Learning (RFML) systems in the future.

Given the natural application of memristor technology to the RF space, it is reasonable to develop threat models within the RF domain and its applications. With this in mind, we develop three threat models in the section below.

\section{Threat Models}

The development of an appropriate threat model is highly context-dependent. In the case of memristor-based analog ML accelerators the context is derived from the edge device manufacturing process, deployment location, and how it interacts with the larger system into which it is integrated. We believe that threat models for memristors should also consider how an adversary might be able to access the information stored in memristors. Access to memristors are split into three categories: White Box (WB), Grey Box (GB) and Black Box (BB). WB access assumes the adversary can provide arbitrary inputs to the model, has prior knowledge about the NN model such as its architecture, and can read the resulting output. Furthermore, an adversary with WB access can read the conductances of individual memristors within the array. BB access assumes the adversary can read input/output pairs but has no further system access. GB access lies between the two, characterizing the adversary as only having partial access to the system. When developing a threat model, it is important to consider who the potential adversaries are and their objectives. For example, the adversary could be a nation-state or commercial competitor. Such an adversary's end goal could be stealing vital technology or inhibiting and sabotaging competition in the market. It then begs the question: What exactly might an adversary try to steal from a memristor enabled device (and why)?

A report by Cottier predicts that the already significant cost associated with training large ML systems will only continue to rise. By 2030, training an AI model could cost as much as \$500 million \cite{cottier_trends_2023}. Reverse engineering others' architectures, hyper-parameters, weights and other details offers a considerably more cost effective path compared to the alternative of developing and training a model independently. Thus, there is a massive economic incentive to steal a NN, including the model's weights. In the case of RF technology, the 6G network currently being developed will likely utilize ML technology for use cases such as network mobility, intelligent routing, energy efficiency, etc. Du et al explains additional avenues in which ML might be implemented in 6G communication technology \cite{du_machine_2020}. Considering the non-negligible role NNs will play in this space, there are significant economic incentives to providing low cost solutions for 6G network infrastructure. As a result, many competitors in the commercial sector have proper economic motivation to steal NN models and their respective weights.

In a military environment, there is not just an economic incentive to need to protect Intellectual Property (IP) from theft, but also a requirement to protect national security and ensure successful missions. In wireless communication, memristors could aid RF signal processing, communication, and signal classification tasks. In addition to the use cases mentioned above in Section \ref{sec:RF Analysis}, memristors can be used to accelerate RFML signal classification such as identification of friend or foe (IFF) or target tracking. Another major concern in wireless technology for the military community is an impersonation of an ally through replay attacks. In recent history, it was reported that Iran was able to capture a US drone by jamming its sensors and then spoofing its GPS coordinates \cite{noauthor_exclusive_nodate}. Unfortunately, memristors have the potential to increase the attack surface of chip technology as compared to its traditional CMOS counterparts; especially since they will be embedded and integrated with CMOS technology. Jamming, spoofing and replay attacks are just a few attack scenarios discussed under the Remote Attack threat model. 

Below, we identify and explore three vital threat models: 1) Supply Chain Attacks, 2) Physical Attacks, and 3) Remote attacks. For each model we discuss potential methods an adversary could utilize to attack memristor-based technology as well as potential defenses. Following this discussion of individual threat models we also explore how vulnerabilities may emerge from a combination of threat models. 

\subsection{Supply Chain Attacks}

Supply chain security is an increasing concern due to the global nature of semiconductor and electronics manufacturing.  Figure 3 shows the global nature of the semiconductor supply chain and vulnerabilities at each stage.  This is a particular concern for military applications since the Department of Defense (DoD) requirements for chips include a wide variety of technology that range from legacy microelectronics (node sizes greater than 100nm) to the current state of the art. There is ``no single manufacturer or manufacturing process than can meet the variety of DoD requirements." Furthermore, said requirements are currently ``dependent on fragile supply chains with a highly concentrated semiconductor industry at center stage" \cite{mondschein_securing_nodate}. Security threats in this category can have WB Access, GB Access, or BB Access depending on the location of attack within the supply chain. The foremost concern in a supply chain threat model is an abundance of counterfeit components flooding the market. 

The presence of counterfeit chips or boards in the supply chain can result in loss in revenue, intellectual property, brand recognition, and reliability in a critical system’s functional capability \cite{arafin_hardware-based_2017}. Various mitigation techniques have been developed for CMOS systems, all of which offer different security properties, deployability and necessary level of effort to integrate into an existing integrated circuit (IC). See \cite{arafin_hardware-based_2017} for an in-depth examination. 

Counterfeits also include the added threat of insertion of hardware Trojans and other malware. ``Hardware Trojans are malicious modifications of a circuit...that aim at manipulating its behavior in an undesired manner." Possible results include denial-of service, change of functionality or the opening of side channels to leak sensitive information. Hardware Trojans are by design difficult to detect via traditional methods including testing of the circuit, optical inspection or side-channel analysis \cite{kumar_parametric_2014}. An adversary can insert a hardware Trojan at any point in the IC supply chain, including during the in house design process, 3rd Party IP vendors, Computer Aided Design (CAD) tools, fabrication, testing, and/or distribution phases. Insertion at the fabrication stage is the most common concern \cite{xue_ten_2020}. In the case of edge analog accelerators such malicious alterations to the circuit could lead to issues of classification error, leakage of sensitive data or device failure. Such issues could appear consistently, intermittently, probabilistically or only when certain conditions are met in the case of a parametric hardware Trojan. A parametric hardware Trojan is ``carefully inserted to modify the electrical characteristics of predetermined transistors in a circuit by altering parameters such as doping concentration and dopant area" \cite{kumar_parametric_2014}. Presently, consideration of hardware Trojans in circuits with memristors has been very limited so far. Initial work has highlighted the increased difficulty in distinguishing between sneak path current and effects due to a stealthy hardware Trojan in 0T1R designs \cite{basu_detection_2021}. Due to the lack of research in this area presently, hardware Trojans currently pose an unabated threat in memristive applications.

\begin{figure}
    \centering
    \includegraphics[scale=0.4]{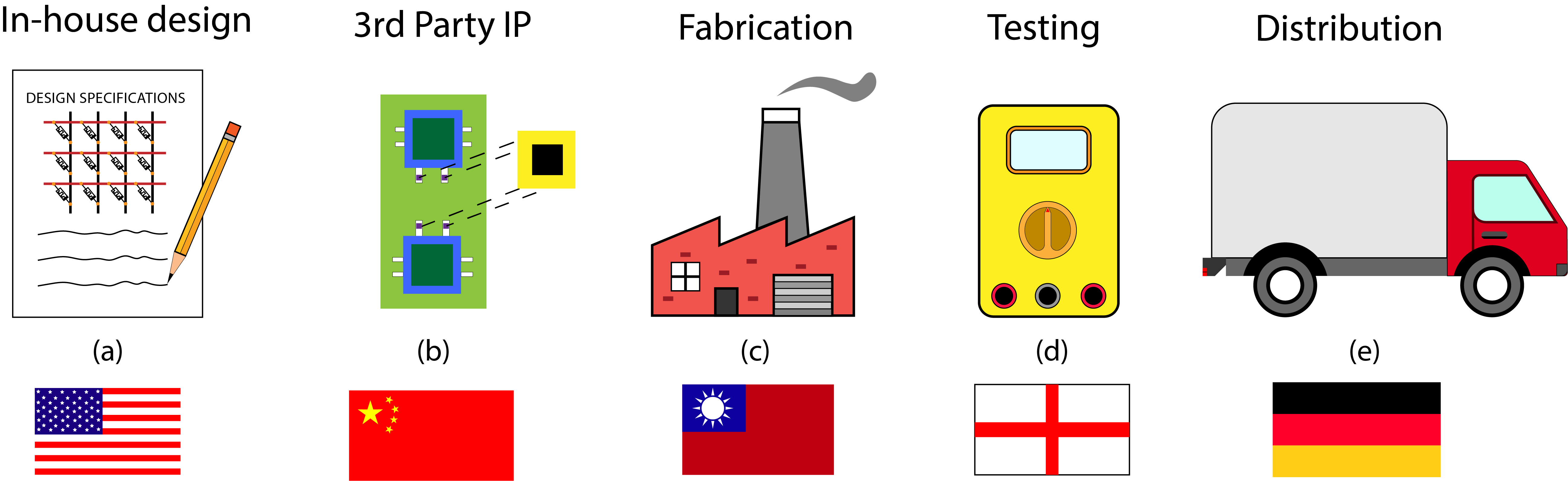}
    \caption[]{Below is list of areas in which an adversary could attack during the supply chain. Note that each step of the supply chain could occur in a different country, which increases the area of the attack surface.\\
    a) Given access to the original design specifications, the adversary could insert side-channels to facilitate leakage of valuable information (USA).\\
    b) Third party products may come from untrustworthy sources, and could contain Trojans or other malicious hardware (ROC).\\
    c) An adversary could produce counterfeit products or compromise the integrity of the chips (TWN).\\
    d) A testing team could reverse engineer the NN model details (e.g. architecture, weights, etc.) by recording the inputs and outputs of the memristor array (ENG).\\
    e) Distributor could illegally retain the chip and/or replace with a counterfeit (DEU).
    }
    \label{fig:Supply_Chain}
\end{figure}

Weight programming introduces an instance of vulnerability made possible by the nature of accessing memristors through its peripherals. Memristor crossbar peripherals, such as the programming/readout ADCs are susceptible to probing. As discussed by Huang et al., the memristor ``array is typically equipped with high resolution ADC for the write/verify scheme to minimize the variations of the cell conductance. Individual cell’s conductance will be read out by such ADC and compared with the reference when initially loading the DNN models on-chip" \cite{huang_new_2020}. Therefore, if an adversary can observe weight programming (e.g. during a manufacturer's testing phase), then they can easily extract memristor weights.  

\subsection{Physical Attacks} \label{Physical}

Gaining physical access to a device opens up numerous attack surfaces to attackers. This translates to GB access to the relevant model in the best case and WB access in the worst case. Assuming a nation state adversary which possesses technological capabilities (such as an advanced lab), then the potential for a large array of attacks that can be carried out is introduced.  Probing allows sensitive data, such as network architecture, proprietary weights or other pieces of sensitive intellectual property to be stolen. 

Huang et al. describes a micro-probing attack in which an adversary with white box access can extract individual weights from a crossbar. Even if the individual cells can't be probed directly, the threat of having their values read out via the crossbar's peripheral circuitry remains \cite{huang_new_2020}. Such an attack would enable an adversary to steal an entire model.

Even under a weaker model where entire memristive crossbars are treated as black boxes (and the probing of individual memristors isn't viable), Wang et al. demonstrates the danger posed by power and time side channel leakage. Treating the tiles as black boxes and without prior knowledge of the network architecture, details such as ``stored layer type, layer sequence, output channel/feature size and convolution kernel size" were able to be reverse engineered \cite{wang_side-channel_2023}.

A fault injection attack could be carried out by altering the network's weights, reducing inference accuracy. Forensic analysis and reverse engineering would grant an adversary valuable information in carrying out other attacks, such as crafting adversarial inputs to an RF classifier \cite{kokalj-filipovic_targeted_2019}.
A captured device also presents numerous opportunities for reverse engineering, from specific circuits and architectures, to recovering stored weights and data \cite{dhavlle_reverse_2022}. 

Model extraction attacks describe the process of an adversary reverse engineering an ML model by collecting many input/output pairs and using those pairs to train a new model. Such attacks pose a threat to all inference engines since such attacks simply require black box access to gather sufficient input-output pairs. This threat has been specifically addressed for memristor-based inference accelerators by Huang et al., who explores the threat as well as memristor-specific countermeasures \cite{huang_new_2020}. This attack is not limited to captured devices since it is possible in remote attacks if the adversary can obtain input/output pairs. 

\subsection{Remote attacks} \label{Remote}

Remote attacks encompass any adversarial action at a distance, where physical access to the device is not available, and are therefore treated as BB access. Previous works \cite{kokalj-filipovic_targeted_2019, roy_machine_2019} have demonstrated the danger of adversarial inputs to ML models in RF applications. The threat of adversarial inputs is not only limited to general misclassification, but also includes the erroneous classification of a specific target class. Jamming attacks enable an adversary in achieving denial-of-service for receiving and/or transmitting of data. All wireless media is vulnerable to jamming due to the shared nature of the wireless medium. Identification of jamming attacks can aid in avoiding or subverting them (e.g. via frequency hopping, reconfigurable antennas or application of cognitive radio). ML techniques have shown promise in accomplishing said identification \cite{jacovic_mitigating_2023}.

As with any application that utilizes RF communications, a further concern is that of impersonation and replay attacks. Countermeasures must be taken in order to prevent an adversary from posing as a trusted entity; whether that be via novel techniques at the memristor/crossbar level or by utilizing techniques higher up the stack (such as Message Authentication Code schemes).

\subsection{Hybrid Attacks}

In addition to threats from the above models, a hybrid attack utilizing multiple categories should be considered. A hybrid attack is more likely to result in successful penetration or denial of service since it combines the attack surfaces available to multiple threat models. For example: an adversary could obtain a device, reverse engineer intimate model information stored on the device (such as weight matrix values) and use that information to launch devastating attacks remotely \cite{kokalj-filipovic_targeted_2019}.  Another example of a hybrid threat is where an adversary inserts a parametric hardware Trojan into the weight matrix (e.g. via a compromised supply chain) and then uses the perturbations to increase the effects of other attacks (e.g. jamming) after the device's deployment.

\section{Countermeasures}

The system of RF/ML/Memristor computation has several interesting security features in addition to the vulnerabilities discussed above.  These include mechanisms for unique identifiers, key generation, true random number generation, lightweight cryptography, and obfuscation of hardware, software, and ML models.  These can all be used as countermeasures to the attacks described above.
Below we describe existing work and open problems in several of these areas.

\subsection{PUFs}

A PUF is a relatively new hardware-based security primitive that ``exploit[s] the...manufacturing variations that occur in almost all physical systems on small length scales." By design, a PUF is ``unclonable, and constitutes an individual fingerprint of each system" \cite{ruhrmair_pufs_2014}. PUFs can serve a variety of purposes, including identification and authentication, key storage, key exchange, digital rights management purposes, tamper detection, and memory encryption \cite{ruhrmair_pufs_2014}. Existing PUFs can be integrated to provide these features to memristive circuits. In particular, PUFs serve as a natural countermeasure to counterfeiting attacks. 

Memristors in various configurations have also been used to construct hybrid-PUFs, promising better performance than their CMOS-only counterparts. Koeberl et al. utilizes a ``weak write mechanism," in which insufficient access time is given to a memristor write operation, leaving the device's conductance in an undefined state determined by device variation \cite{koeberl_memristor_2013}. Gao et al. utilizes the variation of individual memristors within a large crossbar array in combination with ring oscillators to ``translate a memristance value into a frequency through a [current mirror-controlled ring oscillator]" in order to generate random bits \cite{malkin_mrpuf_2015}. Zhang et al. presents a silver (Ag) sputtering technique in which a memristive crossbar serves as a weak PUF, producing a single challenge-response pair \cite{zhang_nanoscale_2018}. Several other weak PUFs designs utilize differential pairs of memristors \cite{jiang_provable_2018, pang_optimization_2017, chen_utilizing_2015}, which compare the conductances of pairs of memristors. Unfortunately, this particular approach requires two (2) cross bars for each bit of key. Moreover, these PUFs do not prevent an adversary from learning the fingerprint through probing from a memristor array, and therefore only provides lightweight security. 

Govindaraj et al. describes an Arbiter PUF design where memristor cells are ``connected in daisy chain fashion selectively between two symmetric paths"\cite{govindaraj_design_2020}. Chatterjee et al. attempts to harden another memristor-based Arbiter PUF design \cite{mathew_novel_2015} against cryptanalysis.\cite{chatterjee_memristor_2016} Uddin et al. leverages experimental measurements of hafnium-oxide memristors, eXclusive OR (XOR)ing, and a column shuffling technique to construct an improved memristor-based Arbiter PUF, designated XbarPUF \cite{uddin_design_2018}. The devices' general lack of availability has limited testing outside of simulation, however, leading to a lack of resistance to modeling attacks in many cases \cite{zeitouni_security_2020}. More recently, Ibrahim et al. actually fabricated a memristor based PUF, allowing for full testing outside of simulation.

Providing security measures for memristor technology that can detect and combat probing or tampering remains a relatively open problem.

\subsection{Lightweight Cryptography}

Traditional cryptographic algorithms and protocols are often ill-suited for resource constrained devices. The recent explosion in IoT devices has led to the development of new lightweight cryptographic protocols for encryption, authentication, integrity checking, and key management. As a result, there are numerous lightweight protocols tailored to various use cases, such as PRINCE, PRESENT, and lightweight variants of traditional algorithms such as Advanced Encryption Standard (AES) and Data Encryption Standard (DES). For an exhaustive listing, see Rana et al \cite{rana_lightweight_2022}.

The usage of memristors to provide efficient crypto-computation in resource constrained settings has been considered, as in Xue et al.'s proposal to integrate area-efficient \textit{in-situ} blockchain technology integrated with a RISC-V processor \cite{xue_risc-v_2019}.There are also lighter weight variants of Public Key Systems such as Elliptic Curves and some Post-Quantum cryptography algorithms \cite{ebrahimi_lightweight_2020}. 

\subsection{Data Flushing}

A potential defense mechanism involves the flushing of onboard data once tampering has been detected. This could be employed to prevent the invasive reading (or overwriting) of proprietary weights in a memristive crossbar array, flush secret keys, cause the flushing of all intermediary values in a VMM operation, etc. Matsuda et al. describes such a flushing mechanism to defend against laser fault injection (LFI) on an AES cryptographic processor. A bulk built-in current sensor (BBICS) provides the chip with 100\% coverage, triggering the flush if a current above a certain threshold is detected in the chip's silicon substrate \cite{matsuda_286_2018}. While improvement is needed, memristor's CMOS-compatibility \cite{wang_resistive_2020} shows promise for the integration of similar detection and/or defense mechanisms.

\subsection{Model Protection Mechanisms}

Deploying Edge NN accelerators in which a model's weights are stored in non-volatile memory poses a heightened risk for various model extraction attacks. There exists a wide range of hardware-agnostic strategies aimed at protecting NN models from unauthorized copying. Lederer et al. provides an extensive exploration of these strategies \cite{lederer_identifying_2023}. Detailing all of these defenses is beyond our scope herein; however, there are a few memristor-centric defenses that we considered, such as \cite{zou_review_2022}. Zou et al.'s survey describes several memristor-specific strategies, including exploiting the memristor's obsolescence effect, model encryption, NN weight transformation, and fingerprint embedding.

The memristor's obsolescence effect describes the gradual degradation of the stored resistance state in an individual device as the result of repeated read voltage pulses. By leveraging this effect, adversaries are prevented from gaining sufficient input/output pairs for a model extraction attack. This strategy requires frequent reprogramming or refreshing of the stored weights, introducing energy overheads, unavoidable delays, and device downtime. 

Model encryption schemes store the NN's weights in an encrypted form, decrypting each time they are needed for inference. Example implementations are presented by Zou et al. and Lin et al.'s respective bias \cite{zou_enhancing_2022} and shuffling \cite{lin_chaotic_2021} techniques. Testing of both strategies was constrained to simulation, however, and did not consider the effect of any potential side channel leakage. In general, this strategy introduces significant latency and energy overheads, both due to the encryption/decryption as well as the frequent reprogramming of the memristor weights. Efforts to optimize this process including limiting encryption to a critical subset of the model's weights have been accomplished \cite{cai_enabling_2020}. However, an adversary with white box access and granular control over the device could access the weights once they are decrypted for use.

The NN weight transformation strategy aims at protecting weights by altering how they are stored on the physical device. Existing approaches include obfuscating row and/or column connections either within the same or multiple crossbars. 

The fingerprint embedding strategy embeds device-specific information into the programmed NN weights, rendering them useless if transplanted onto another device. Huang et al. explored the use of ADC offsets for this defense strategy \cite{huang_new_2020}.

\subsection{Design Obfuscation}

Design Obfuscation is a technique which shields the hardware details against acquisition by an unauthorized party. This can be used to protect IP, but also to defend against reverse engineering and higher level attacks. In addition to demonstrating the danger of power and time side channel leakage, Wang et al. also proposes several obfuscation-based countermeasures to side channel attacks. These countermeasures include the scrambling of crossbar tile starting times, dummy inputs to peripheral circuitry components (such as ADCs) and power equalization schemes \cite{wang_side-channel_2023}. Huang et al. \cite{huang_new_2020} presents a input shuffling scheme that defends against an input/output pair attack described in section Section \ref{Physical}. 

\subsubsection*{Logic Locking}

Logic Locking is a defense mechanism developed in response to threats made prominent by the globalization of the integrated circuit supply chain. The new era of fabless semiconductor companies requires interaction with multiple third party (potentially untrusted) entities. The threats posed by such a distributed manufacturing process include intellectual property piracy, reverse engineering, and hardware Trojans. Logic locking attempts to add additional security to integrated circuit designs by inserting additional logic into a circuit and locking the original design with a secret key \cite{yasin_evolution_2017}. Jiang et al. propose a logic locking/unlocking scheme with provable key destruction solely with a memristor crossbar array. Such an approach ``enables the same crossbar to be used for both security and computing/memory applications, saving chip space while increasing power efficiency" \cite{jiang_provable_2018}.

\subsection{Watermarking \& Stenography}

Watermarking of computation and data provides another security capability.  In contrast to the classic Confidentiality-Authenticity-Integrity,  watermarking hides an identifier into a set of data such that it can not be detected, removed or altered. In this case, output data from a protected ML system can be used illegally by an adversary, however the provenance of the data can be detected either at run-time or in a forensic analysis through the use of watermark inserted during the ML inference computation \cite{boenisch_systematic_2021}. A watermark should also be robust to modifications of the data such as cropping, compression and other manipulations \cite{boenisch_systematic_2021}.  Figure 4 shows the basic idea behind adding a watermark to detect the origin of a particular model output.

The RFML identification problem clearly can benefit from watermarking to protect model IP, original data, and proprietary algorithms (as well as for forensics).  Furthermore, it should be possible to efficiently implement current watermarking algorithms in memristor technology by modifying the stored weights.  Co-locating the watermarking with the computation also reduces vulnerabilities. Watermarking can be implemented on the weights or the input/output data.  Challenges include implementing the watermarking computation in a fully-analog (multi-level) memristor crossbar as opposed to using only binary nodes \cite{chen_-line_2021}. The inherently imprecise nature of memristors will add additional challenges, such as distinguishing a legitimate watermark (or lack thereof) from benign noise or modification by a sophisticated adversary.

Watermarking has been widely implemented with image-based media. Although there are a few image watermarking methods applied to memristors as presented in \cite{sehra_secure_nodate}, there has not been extensive research performed on watermarking memristor weights in general, especially those used in an RF-based context. 

The ability to watermark a model's output data could be leveraged to prevent impersonation, replay, or other similar attacks. Chang and Su present a NN-based watermarking scheme in which an input image is fed into a neural network and a watermarked copy of the image is then outputted \cite{chuan-yu_chang_neural-network-based_2005}. Such a scheme could conceivably be integrated into an existing system by feeding the output neurons of the main network into the inputs of the watermarking network. Huang et al. present a similar scheme in which the watermark could be recovered from a watermarked image without the original image or watermark information known \textit{a priori} \cite{song_huang_blind_2008}.

\begin{figure}
    \centering
    \includegraphics[scale=1.135]{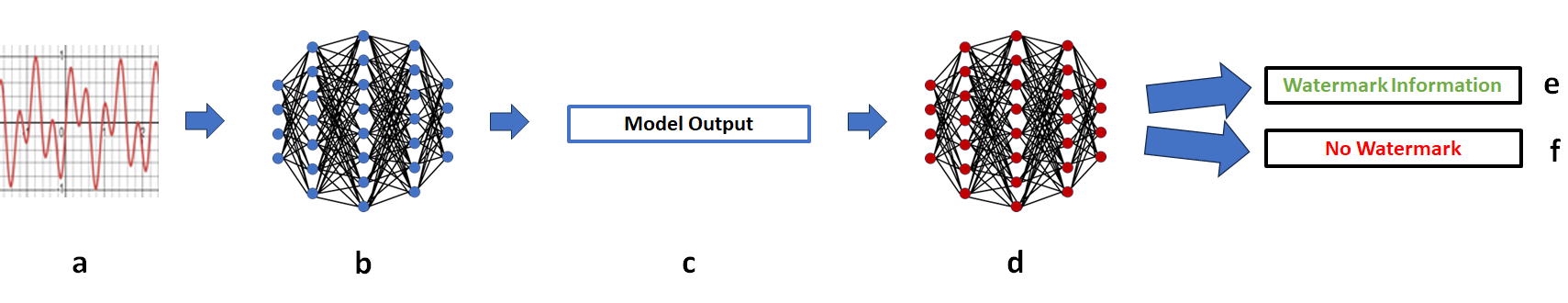}
    \caption[]{A flow chart depicting the potential use of watermarking to detect imposter/ compromised classifiers. An RF signal (\textbf{a}) is received by an inference accelerator (\textbf{b)}, which then performs a computation on the signal and sends its output. The output is received by a watermark extractor (\textbf{d}), which either detects watermarked information in the output (\textbf{e}) or detects the lack of a watermark (\textbf{f}) indicating a compromised device.
    }
    \label{fig:Watermarking}
\end{figure}

\subsection{Anomaly Detection \& Fault Tolerance}

Anomalies, whether the result of an adversary or benign, can be detected in a variety of sensors including temperature, voltage, current \cite{matsuda_286_2018}, and signal integrity. In addition to security concerns, the information gathered by such sensors can be used to improve device performance. Zhao et al. demonstrates this by deploying a monitoring infrastructure of sensors to effectively implement dynamic frequency scaling on multicore processors \cite{zhao_dedicated_2011}.

Due to an immature fabrication process, as well as various device non-idealties, the transfer of a NN model onto a memristive crossbar isn't a straightforward task. However, NNs (and in particular DNNs) are generally known to be fault tolerant. This useful property is also observed when such architectures are deployed on memristive accelerators, despite various device-related defects and variation. This tolerance doesn't extend indefinitely, however, as there are typically a non-trivial number of critical faults (CFs) that can lead to inference errors, including misclassification \cite{chen_pruning_2021}. Faults due to device non-idealties, manufacturing variations, and failures are of particular interest. Examination of fully hardware memristive NNs shows that small nonlinearity affects accuracy minimally, but large non-linearities can considerably degrade the model's performance \cite{kiani_fully_2021}. The same model also demonstrated high tolerance for variance in the hardware-based ReLU's gain. 

Sun and Yu examine the effects of memristor non-idealities on both training and inference via simulation, assessing the non-linearity and asymmetry of conductance tuning, manufacturing variations, endurance, and retention. Following this examination, the following conclusions were reached:

\begin{itemize}
    \item Training accuracy is more sensitive to the asymmetry of conductance tuning than the nonlinearity.
    \item Conductance range variation does not degrade the training accuracy, instead, a small variation can even reduce the accuracy loss introduced by asymmetry.
    \item Device-to-device variation can also remedy the accuracy loss due to asymmetry while cycle-to-cycle variation leads to significant accuracy degradation.
    \item Different drifting modes affect the inference accuracy differently. The best case is where the conductance is drifting up/down randomly, rather than systematically in one direction \cite{sun_impact_2019}.
\end{itemize}

Several ML-based techniques have been developed to identify and prune CFs from the network. The removal of said CFs reduces the number of redundant columns necessary for effective fault tolerance by up to 99\% \cite{chen_pruning_2021, chen_efficient_2022}.

Online fault detection is of particular interest for the crossbar arrays, as both adversarial action and material aging effects can lead to a loss of the programmed value for weights in the network. Several methods for online fault detection have been developed to solve this problem, including Chen and Chakrabarty's method that utilizes a watermarking-based backdoor schema \cite{chen_-line_2021}. In this schema, the neural network can be checked for faults by periodically attaching watermarks to otherwise normal inputs. Due to the model being over-fitted to the watermarks beforehand, faults have a high likelihood of causing inference error on said inputs. Unfortunately, present NN watermarking has largely been relegated to image data so far \cite{boenisch_systematic_2021}. Research in the area hasn't extended to the RF, or other non-image domains. More general and efficient on-line fault detection techniques are in need of development. Anomalies can also be detected through specialized sensor chiplets or package-level sensors in chiplet architectures.

\section{Package Security using Chiplets}

\begin{figure}
    \centering
    \includegraphics[scale=0.3]{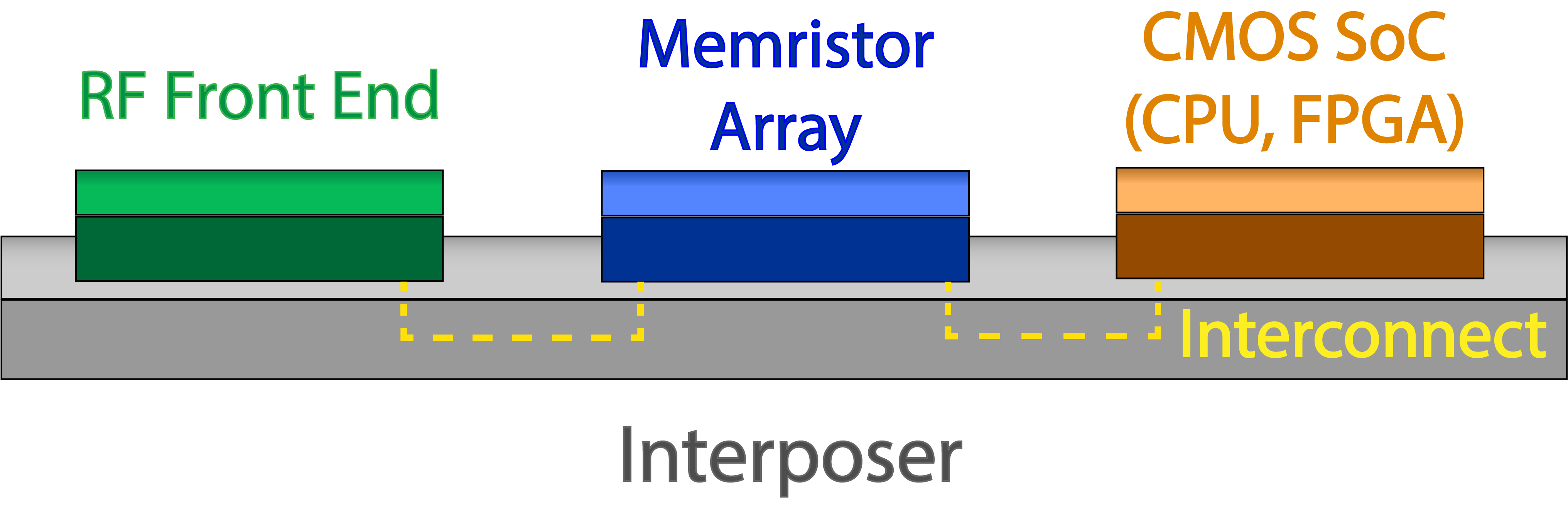}
    \caption{A sample chiplet architecture, integrating RF receivers/ transmitters, one or more memristive crossbars for analog signal processing, and a CMOS System on Chip (SoC) to handle general compute.  These are integrated on a common silicon interposer that provides high speed interconnects between the chiplets as well external interfaces, power delivery and potentially active circuits.}
    \label{fig:Memristor_Chiplet}
\end{figure}

While memristors offer tremendous benefits in certain applications, they do not offer the same general purpose capabilities of typical CMOS systems. In order to realize the benefits of a memristive accelerator in a non-laboratory environment, this technology should be integrated with other more traditional computing paradigms. Current research commonly utilizes custom-built peripheral circuitry to accomplish this task \cite{huang_time-encoding_nodate, kiani_fully_2021, li_efficient_2018}. For a more tightly integrated system in package, a chiplet architecture is a strong candidate instead of a monolithic system. A sample architecture for RF applications is shown in Fig. \ref{fig:Memristor_Chiplet}.

We now discuss vulnerabilities and countermeasures that are specific to chiplet-based systems, referring to the threat models developed in Section V.

\subsection{Chiplet-based Vulnerabilities}

\begin{figure}
    \centering
    \includegraphics[scale=0.3]{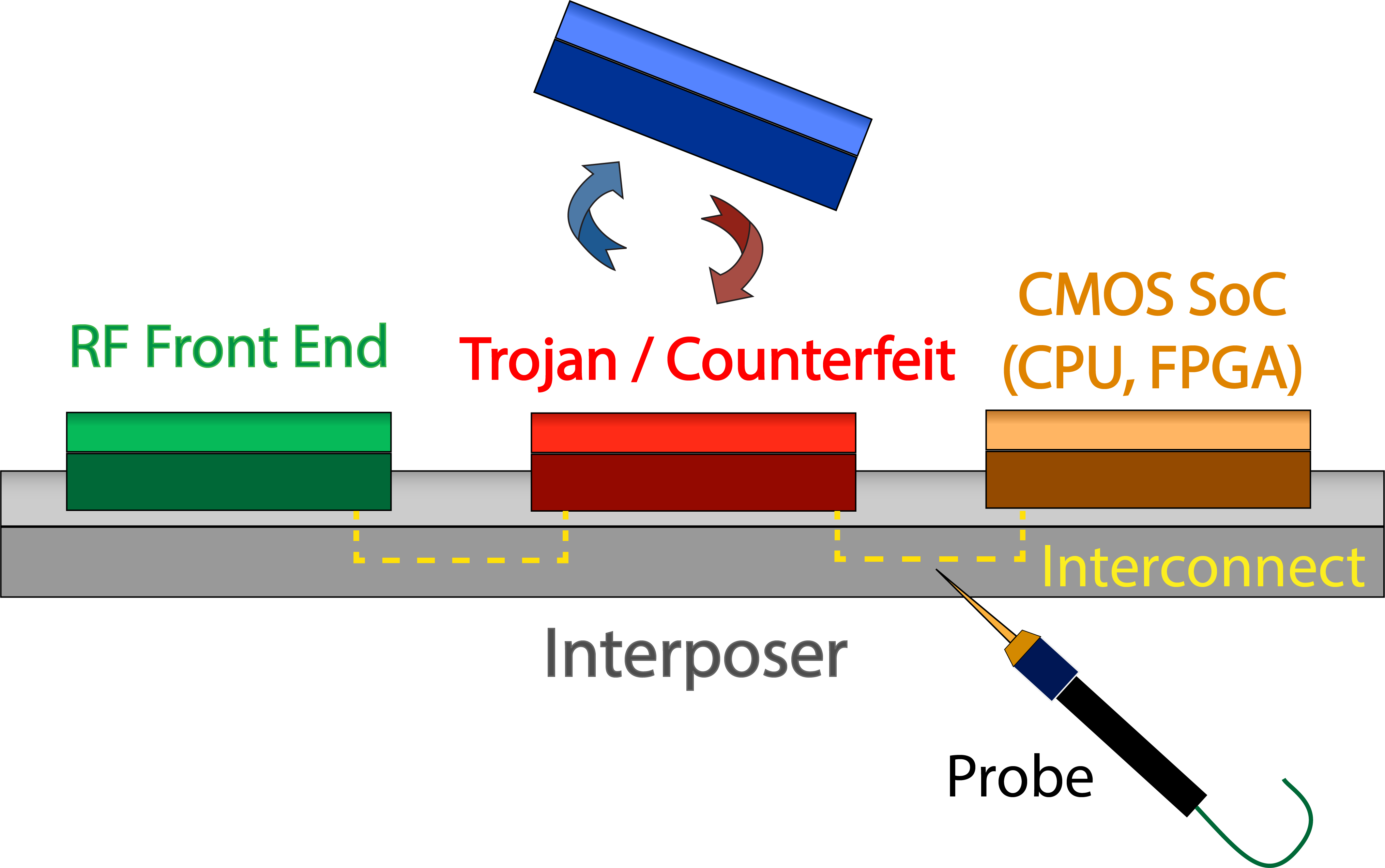}
    \caption{The figure above shows two potential attacks against a chiplet type architecture: Die-swapping, and probing of the interconnections between chiplets.}
    \label{fig:Die_Swapping}
\end{figure}

With the benefits of chiplet designs \cite{httpswwwfacebookcom48576411181_3_nodate} comes new security risks that must be addressed, both general and specific to memristors. Concerns include the threat of die swapping (Fig. \ref{fig:Die_Swapping}), tampering, and the increased ease of invasive probing of the die interconnects. Chiplet security issues fall under both the Supply Chain and Physical Attack threat models.

Counterfeits are a steadily rising threat in the global supply chain. For example, according the European Union Intellectual Property Office (EUIPO) Crime Tech Report \cite{noauthor_intellectual_2022}, nearly one out of five smartphones sold are counterfeit. Moreover, the loss in global sales due to counterfeit devices averaged to approximately 13\%. EUIPO also observed that some regions, such as Latin America, incurred almost 20\% in sales losses. In a chiplet model, many of the components will likely be commercial off-the-shelf (COTS), and sourcing these parts can be risky without an efficient method to validate authenticity. 

Hardware Trojans from untrustworthy chiplets may employ side channel analysis attacks as demonstrated in \cite{lin_moles_2009} in order to obtain valuable information. Chiplets are especially susceptible to this type of attack because the Power Delivery Network (PDN) is shared amongst all of the chiplets \cite{kim_chipletinterposer_2021}.

Chiplet architectures are also vulnerable to direct or EM probing since the inter-die wires are larger and more accessible than on-die wires.

\subsection{Chiplet Security Countermeasures}

One methodology that validates chiplets is proposed by Vashistha et al. \cite{vashistha_trust_2022}. Their methodology suggests using a scanning electronic microscope (SEM) to physically measure, capture, and catalog internal component images and layouts into a database. Although their methodology claims great effectiveness in dealing with tampering, Trojans, and counterfeits, SEM methods are known to be very cost- and time-intensive (sometimes requiring several hours to inspect a single component). Furthermore, decapsulation of the component is often necessary. These downsides will likely prove prohibitive for inspections of volumes of even moderate size. 

Other points of adversarial access include probing, decapsulation, delidding, and polishing. Mosavirik et al. demonstrated that nearly any sort of probing, polishing, or merely even scratching the surface of a chiplet will alter the impedance in the PDN \cite{mosavirik_impedanceverif_2022}. Mosavirik et al. takes advantage of this phenomena by developing a method of authentication which can detect probing, tampering, and hardware Trojans.

Another countermeasure that presents a novel method to authenticate chiplet products is presented by Deric and Holcomb. They propose developing a PUF through delay measurement across the interconnects between chiplets \cite{deric_know_nodate}. This authentication strategy could be used as an effective PUF and is more time efficient compared to \cite{vashistha_trust_2022} above. While this approach is promising, tampering and probing is not examined in detail therein; thus additional exploration to verify PUF robustness is called for. 

An industry perspective on chiplet security is presented in \cite{sperling_security_2022}. Many of the security challenges and techniques from SoC design apply to Chiplet-based design.  In particular the assembly of and communication between multiple untrusted IP cores, and the protection of the IP in those cores, are similar in SoC and Chiplet-based design.  Differences include unintended use models of chiplets and different business models.
Solutions include the ``activation" of chiplets through certificates and the use of block-chain ledgers and other cryptographic techniques \cite{xu_electronics_2019}.

Circuit and software obfuscation can be combined with techniques to protect IP and prevent reverse engineering.
Finally, standards are being developed which include security aspects such as Universal Chiplet Interconnect Express (UCIe) \cite{das_sharma_universal_nodate}, the Open Domain-Specific Architecture (ODSA) \cite{drucker_open_2020}, and Compute Express Link (CXL)\cite{das_sharma_compute_nodate}, as well network-on-chip (NoC) approaches.

\section{Open Problems}

There are numerous open problems in the combination of  RF/ML/memristor security. They range from RF datasets and threat models, to ML vulnerabilities, to the memristor computational fabric and its packaging into an overall system.  Novel attacks and defenses can be envisioned at many levels and across technologies.  In many cases, existing work can be leveraged and modified to fit the application and computational assumptions.

A few open problems that have been mentioned earlier include:

\begin{itemize}
    \item Potential hardware Trojans in memristor technology as well as methods for detection and prevention.
    \item The role of approximate computing in terms of both vulnerabilities and defense \cite{yellu_security_2021}.
    \item Additional research into efficient in-situ fault detection and recovery for memristor arrays \cite{chen_-line_2021}.
    \item More advanced memristor PUF designs, with particular focus on resilience to modeling attacks \cite{zeitouni_security_2020, hutchison_bitline_2014}.
    \item For each threat model (Supply Chain Attacks, Physical Attacks, and Remote Attacks), additional vulnerabilities and defenses are identified, emphasizing the RF applications.
    \item Development of watermarking and design obfuscation techniques that can integrate CMOS with memristor crossbar technology.
    \item Producing reliable and effective PUFs and TRNGs that utilize memristor architectures.
    \item A deeper investigation of security vulnerabilities relevant to memristor chiplet architectures.
    \item Developing more precise threat models for each of the three categories 1) Supply Chain, 2) Physical Attacks, 3) Remote Attacks. 
    \item Developing on-chip sensors and front-end processing technology specific to memristor crossbar arrays designed for a zero-trust environment.
    \item Thorough physical testing of memristor-based concepts (e.g. PUFs, fault tolerance schemes, etc.) that have previously been verified only by simulation.
\end{itemize}

\section{Conclusion}
Memristors have the potential to accelerate RF signal processing in both energy efficiency and speed. With these benefits comes the potential for new vulnerabilities, many originating from the device's ability to store information in a non-volatile manner. With the exorbitant costs required to develop ML models, competitors have every incentive to steal the model architecture and weights. We presented a broad survey on security for memristor architecture with respect to the RF domain. We organized potential adversarial action into three threat models: Supply Chain Attacks, Physical Attacks, and Remote attacks. Next we described potential defenses and countermeasures for attacks under each of the threat models. We also provided a brief review of security for chiplet architectures since this is a likely format in which memristor technology will be integrated and deployed. Finally, we defined several open problems and potential research directions that should serve to mitigate vulnerabilities and foster robust implementations of memristor technology.

\section*{Acknowledgment}

Thanks to our fellow researchers Professor Qiangfei Xia, Ali Abdel-Maksoud, and Yi Huang for their technical support. 

\AtNextBibliography{\footnotesize} 
\printbibliography

\end{document}